\documentclass[letter]{ieice}

\usepackage[dvips]{graphicx}

\field{}
\title{Choice Disjunctive Queries in Logic Programming}

\authorlist{
 \authorentry{Keehang KWON}{m}{labelA}
\authorentry{Daeseong KANG}{n}{labelB}
}
\affiliate[labelA]{The author is a professor of Computer Eng., DongA University. khkwon@dau.ac.kr}
\affiliate[labelB]{The author is a professor of Electronics Eng., 
DongA University.}

\received{2003}{1}{1}
\revised{2003}{1}{1}
\finalreceived{2003}{1}{1}

\newenvironment{numberedlist}
{\begin{list}{\makebox[20pt]{\hss(\arabic{itemno})\enspace}}
             {\usecounter{itemno}\labelwidth 20pt}}{\end{list}}

\newcounter{itemno}

\newcounter{itemno1}

\newcounter{itemno2}
\newcounter{lemma}
\newcounter{exno}

\newcounter{defno}

\newenvironment{defn}{\refstepcounter{defno}\medskip \noindent {\bf
Definition \thedefno.\ }}{\medskip}

\newcommand{\sep}{\;\vert\;}

\newcommand{\oprove}{\vdash\kern-.6em\lower.7ex\hbox{$\scriptstyle O$}\,}

\newcommand{\Pscr}{{\cal P}}

\newcommand{\pderivation}{{\cal P}\kern -.1em\hbox{\rm -derivation}}
\newcommand{\pderivationl}{{\cal P}\kern -.1em\hbox{\em -derivation}}
\newcommand{\pderivable}{{\cal P}\kern -.1em\hbox{\rm -derivable}}
\newcommand{\pderivablel}{{\cal P}\kern -.1em\hbox{\em -derivable}}
\newcommand{\pderivations}{{\cal P}\kern -.1em\hbox{\rm -derivations}}
\newcommand{\pderivability}{{\cal P}\kern -.1em\hbox{\rm -derivability}}

\newcommand{\all}{\forall}
\newcommand{\some}{\exists}

\newcommand{\ie}{{\em i.e.}}

\newsavebox{\lpartfig}
\newsavebox{\rpartfig}

\newenvironment{exmple}{
 \begingroup \begin{tabbing} \hspace{2em}\= \hspace{3em}\= \hspace{3em}\=
\hspace{3em}\= \hspace{3em}\= \hspace{3em}\= \kill}{
 \end{tabbing}\endgroup}

\newcommand{\lb}{\langle}
\newcommand{\rb}{\rangle}
\newcommand{\pr}{prov}

\newcommand{\prove}{exec} 
  
\newcommand{\add}{\oplus} 

\newtheorem{theorem}[lemma]{Theorem}

     {\\* \hspace*{\fill} \end{trivlist}}

\newcommand{\muprolog}{{Prolog$^{\add}$}}
\renewcommand{\pr}{pv}
\renewcommand{\prove}{ex} 
\newcommand{\defeq}{:-}
\renewcommand{\Pscr}{{\cal P}}

\begin{document}
\maketitle
\begin{summary}
One of the long-standing research problems on logic programming is to treat the cut predicate
 in a logical, high-level way.
 We argue that this problem can be solved by adopting linear logic and 
   choice-disjunctive goal formulas of the form 
 $G_0 \add G_1$  where  $G_0, G_1$ are
goals.   These goals have the following intended semantics:
 $choose$ the true disjunct $G_i$ and execute $G_i$ where $i (= 0\ {\rm or}\ 1)$, while $discarding$
 the unchosen disjunct.  
Note that only one goal can remain alive during execution. These goals thus allow us to specify
mutually exclusive tasks in a high-level way.   Note that there is another use of cut which is for breaking out of failure-driven loops and efficient heap management.
Unfortunately, it is not possible to replace cut of this kind with use of choice-disjunctive goals.

\end{summary}
\begin{keywords}
Game semantics; cut; mutual exclusion
\end{keywords}




\section{Introduction}

One of the long-standing research problems on logic programming is to treat the extra-logical 
primitive in a  {\it high-level, logical} way.
The advances of logic programming -- especially structured proof theory --  have enriched Horn clauses with
 additional programming primitives  in a  high-level way
 (higher-order programming, modules, local constants, etc) \cite{MNPS91,Kwon,Loke}.
 Nevertheless some key constructs could not be dealt with in a high-level way, in particular when we are
 concerned with  mutual exclusion (and  the 
cut
predicate). 

Consequently, much attention \cite{Por,Sau08,Kri11} has been given to finding a 
semantics that captures the cut predicate.
These proposals  - based on such  notions of $if$-$then$-$else$ and $until$ -- are quite interesting but somewhat complicated
than necessary.

    In this paper, inspired by the work in \cite{Jap03,Jap08,JapCL1,JapCL2,JapCL12}, we propose a purely logical solution to this problem. It involves the direct employment of linear logic
\cite{Gir87}
 to allow for choice-disjunctive   goals.
 A  choice-disjunctive  goal  is of the form 
$G_0 \add  G_1$ 
 where $G_0, G_1$ are goals. (A more intuitive name would be $choose(G_0,G_1)$.)
Executing this goal with respect to a program $\Pscr$ -- $\prove(\Pscr,G_0 \add  G_1)$ --
 has the following intended semantics: 

\[  \mbox{\rm choose a true  one  between}\ \prove(\Pscr, G_0), \prove(\Pscr, G_1). \]

\noindent For example, given a program $\{ male(kim), female(lee) \}$, execution will succeed on
the goal $male(kim) \add female(kim)$ by choosing $male(kim)$. Similarly,  execution will succeed on
the goal $male(lee) \add female(lee)$ by choosing $female(lee)$. On the other hand,
consider a goal $male(kim) \add female(lee)$. In this case, both disjuncts can lead to a success, and,
in our semantics, it does not matter which disjunct to use. For simplicity, we assume from now on 
that execution always chooses the first
successful disjunct if there are many. Thus, back to the above, execution will succeed on
the goal $male(kim) \add female(lee)$ by choosing $male(kim)$.
Note that the class of choice disjunctive goals is a superset of 
 the class of mutually exclusive goals.

    Another
illustration of this construct is provided by the following definition of the
relation $son(X,Y)$ which holds if $Y$ is a son of $X$.:

\begin{exmple}
$      son(X,Y)$ ${\rm :-}$ \> \hspace{6em}      $(male(X) \otimes father(Y,X))\ \add$\\ 
 \> \hspace{6em} $(female(X) \otimes mother(Y,X)).$
\end{exmple}
\noindent
The body of the definition above contains a mutually exclusive goal, denoted by $\add$.
 As a particular example, solving the query $son(tom,Y)$ would result in selecting and 
executing the first goal $male(tom) \otimes father(tom,Y)$, while discarding the second one.  The given goal will
succeed, producing solutions for $Y$. Of course, we can specify mutually exclusive goals using cut
in Prolog, but it is well-known that cuts
complicates the declarative meaning of the program. Our language makes it possible to formulate mutually exclusive goals in a high-level way.
The class of choice disjunctive goals is, in a sense, a high-level abstraction for the cut
predicate.

     As seen from the example above, choice-disjunctive  goals can be used to perform mutually exclusive
 tasks.
There are several well-designed linear logic languages \cite{HM94,Wini} 
 in which
goals of the form $G_0 \oplus G_1$ are present. A common 
problem of these works  is their treatment of the 
$\oplus$-goals: these goals are treated as inclusive-OR (or classical disjunctive) goals rather than  
exclusive-OR ones:

\[ \prove(\Pscr,G_0\add G_1)\ {\rm if}\  
\prove(\Pscr, G_0)\ \lor \prove(\Pscr, G_1) \]
\noindent where $\lor$ represents classical disjunction. 
Hence, it is rather unfortunate that the declarative reading of $\oplus$ -- 
known as {\it the machine's choice} --
 is missing in these languages.

A  satisfactory solution can be obtained by adopting game semantics of \cite{Jap03}, \ie, by
adding an extra layer of the choice action, as discussed above,
 to their execution model of $\add$. In this way, 
 the execution respects the declarative
reading of $\oplus$, while maintaining provability.
Hence, the main difference is that, once a goal is chosen,
the unchosen goal will be discarded in our language, while it will remain alive (typically through a
creation of a choicepoint) in those languages.

\section{Reconsidering the Foundation in Logic Programming}\label{sec:intro}

 The computation-as-deduction approach\cite{MNPS91,Miller21} has provided a basis for logic programming.
It views the state of a computation as a sequent  and  computing as the  proof search. 
This approach has proven useful, leading to several extensions.
The first such extensions to the Horn clause include
 hypothetical and universally quantified goal formulas, pioneered in \cite{MNPS91}.
 Additional extensions were made using higher-order quantification and linear
logic\cite{HM94}.

Unfortunately,  this approach is appropriate only for  computation with boolean
semantics, i.e., deciding where some formula is true/false. However, this view is too limiting. 
Instead, we believe that computation should be based on a bigger paradigm, \ie, task/game semantics\cite{Jap03,Japtow}.
That is, computation should be interested in deciding whether some formula can be {\it made} true or not.
From this viewpoint, the major criterion for judging the adequacy of a 
logic programming can be explained as follows: 

\begin{itemize}

\item The first phase -- the proof phase -- should be sound and complete with respect to the given semantics such as intuitionstic, classical logic
 or linear logic.

\item The second phase -- the execution phase -- should respect the declarative readings of 
logical connectives.  

\end{itemize}
\noindent  
 Consider, for instance,  $P \otimes Q$ in a query. It reads as follows: solve $P$ and $Q$ concurrently.
 Declarative readings of other connectives are given in the next section.

The sequent
calculus for Prolog  does not violate a correspondence between the declarative
meaning of  logical connectives and proof search operations. Unfortunately, there is no such
guarantee for new connectives.  
$\add$  is such an example where the logical connective  and the proof search operations do not  correspond.
 
\section{Declarative reading of logical connectives}\label{s2}

In this section, based on \cite{HM94,Jap03},  an   overview of declarative reading of linear logical connectives is given
in the sense of  intuitionistic linear logic.

\begin{description}

\item[Additive operations]

The choice group of operations:   $\add$, $\some x$ (and negative occurrences of $\all x$) are defined below.

$A_0\add A_1$ is the problem where, in the initial position, only the machine has a legal move which consists in 
choosing a value 0 or 1. After the machine makes a move $c\in\{0,1 \}$, 
the problem becomes  $A_c$. 

$\some xA(x)$ (and negative occurrences of $\all xA(x)$) is the following: the machine must choose a value $v$ for $x$ and the problem becomes $A(v)$. 

\item[Muliplicative operations]

Playing $A_0\otimes A_1$ means solving the two problems concurrently.  In order to succeed,  
the machine needs to solve each of two problems. 

\item[Reduction]

 $A\supset  B$ is the problem of reducing $B$ ({\em consequent}) to $A$ ({\em antecedent}).  

\end{description}

\section{\muprolog\ with the Old Semantics}

The language is a version of Horn clauses
 with choice-disjunctive goals. It is also a subset of Lolli\cite{HM94}.
 Note that  we disallow linear clauses here, thus allowing only
reusable clauses. 
It is described
by $G$- and $D$-formulas given by the syntax rules below:
\begin{exmple}
\>$G ::=$ \>   $\top \sep A \sep t = s \sep  G \otimes  G \sep    \some x\ G \sep  G \add G $ \\   \\
\>$D ::=$ \>  $A  \sep G \supset A\ \sep \all x\ D $\\
\end{exmple}
\noindent
\newcommand{\sync}{up}
\newcommand{\async}{down}

In the rules above,  $t, s$  represent  terms, and  
$A$  represents an atomic formula.
A $D$-formula  is called a  Horn
 clause with choice-disjunctive goals. A set of $D$-formulas is called a program.

 We will  present a machine's strategy for this language given in \cite{HM94}.
These rules in fact depend on the top-level 
constructor in the expression,  a property known as
uniform provability\cite{MNPS91,KK07}. 
 Note that execution  alternates between 
two phases: the goal-reduction phase 
and the backchaining phase. 
In  the goal-reduction phase, the machine tries to decompose a goal $G$.
If $G$ becomes an atom, the machine switches to the backchaining mode. 
This is encoded in the rule (2).

\begin{defn}\label{def:semantics}
Let $\sigma$ be an answer substitution and let $G$ be a goal and let $\Pscr$ be a set of
$D$-formulas.
Then the task of proving $G$ with respect to $\sigma,\Pscr$ -- 
$\pr(\sigma, \Pscr\vdash G)$ -- 
is defined as 
follows:

\begin{numberedlist}


\item    $\pr(\sigma, \Pscr\vdash \top)$. \% success

\item    $\pr(\sigma, \Pscr\vdash t=s)$ if $t$ and $s$ are unifiable. 

\item $\pr(\sigma, \Pscr\vdash A)$ if $A'\ \defeq\ B\in\Pscr$ and  $A'\theta = A\sigma$ and
$\pr(\sigma\theta, \Pscr\vdash B)$. \% DefR (backchaining)

\item $\pr(\sigma, \Pscr\vdash G_0 \otimes G_1)$  if $\pr(\sigma, \Pscr\vdash G_0)$ and 
$\pr(\sigma, \Pscr\vdash G_1)$. 

\item $\pr(\sigma, \Pscr\vdash G_0 \add G_1)$  if $\pr(\sigma, \Pscr\vdash G_0)$.

\item $\pr(\sigma, \Pscr\vdash G_0 \add G_1)$  if $\pr(\sigma, \Pscr\vdash G_1)$.

\item $\pr(\sigma, \Pscr\vdash \some x G)$  if $\pr(\sigma\sigma_1, \Pscr\vdash\ [w/x]G)$ 
where $w$ is a new free variable, $\sigma_1 =  \{\lb w,t\rb \}$ and $t$ is a term.


\end{numberedlist}
\end{defn}
\noindent 
Initially, $\sigma$ is an empty substitution. $\sigma\theta$ in Rule 3 represents the composition of  two substitutions $\sigma$ and $\theta$. In the above, most rules are straightforward to read.

As an   
illustration of this approach, let us consider the following program $\Pscr$. \\

$ \{\ emp(tom)\ \defeq\ \top.\ emp(pete)\ \defeq\ \top.\ $

$   harvard(tom)\ \defeq\ \top.\ mit(pete)\ \defeq\ \top.\ \} $ \\

Now, consider a goal task 
 $\some x ( (yale(x) \oplus harvard(x)) \land   emp(x))$.
 
The following is a proof tree of this example.  \\

$\{ (w_0,tom) \}$, $\Pscr \vdash \top$ \% success

$\{ (w_0,tom) \}$, $\Pscr \vdash harvard(w_0)$ \% defR

$\{ (w_0,tom) \}$, $\Pscr \vdash  yale(w_0) \add harvard(w_0)$ \% $\add$-R

$\{ (w_0,tom) \}$, $\Pscr \vdash \top$ \% success

$\{ (w_0,tom) \}$,  $\Pscr \vdash emp(w_0)$ \% defR

$\{ (w_0,tom) \}$, $\Pscr \vdash ( (yale(w_0) \add harvard(w_0)) \land  emp(w_0))$ \% $\land$-R

$\emptyset$, $\Pscr \vdash \some x  ( (yale(x) \add harvard(x)) \land  emp(x))$ \% $\some$-R \\

The following theorem connects our language to linear logic.
Its proof is easily obtained from \cite{HM94}.

\begin{theorem}
 Let ${\cal P}$ be a program and 
let $G$ be a goal. Then, $\pr(\Pscr,G)$ terminates with a success
 if and only if $G$ follows from
$\Pscr$ in intuitionistic linear logic. Furthermore, the interpreter respects the declarative
reading of all the logical connectives except $\add$.
\end{theorem}

\section{The Execution Phase}\label{sec:0627}

Adding game semantics  requires another execution phase beside the proof phase.
To be precise, our new execution model -- adapted from \cite{Jap03} -- 
 actually solves the goal relative to the program using the proof tree built in the proof
phase.

 In the execution phase, it just follows the path in the proof tree, printing the output values
occasionally.

\begin{numberedlist}


\item    $\prove(\sigma, \Pscr\vdash \top)$. \% success

\item    $\prove(\sigma, \Pscr\vdash t=s)$ if $unify(t,s)$. \% invoke unification.

\item $\prove(\sigma, \Pscr\vdash A)$ if 
$\prove(\sigma\theta, \Pscr\vdash B)$, provided that the former is derived from the latter via $DefR$. 

\item $\prove(\sigma, \Pscr\vdash G_0 \otimes G_1)$  if $\prove(\sigma, \Pscr\vdash G_0)$ and 
$\prove(\sigma, \Pscr\vdash G_1)$, provided that the former is derived from the latter via $\land$-R.

\item $\prove(\sigma, \Pscr\vdash G_0 \add G_1)$  if $\prove(\sigma, \Pscr\vdash G_i)$ where $i$ is 0 or 1, provided that the former
is derived from the latter via $\add$-R.

\item $\prove(\sigma, \Pscr\vdash \some x G)$  if $\prove(\sigma\sigma_1, \Pscr\vdash\ [w/x]G)$, 
provided that the former
is derived from the latter via $\some$-R.


\end{numberedlist}
\noindent 
Note that only the $\add$-R rule has changed.

The following theorem  justifies our machine with respect to linear logic.
Its proof is easily obtained from the analysis of the above algorithm.

\begin{theorem}
 Let ${\cal P}$ be a program and 
let $G$ be a goal. Then, $\prove(\Pscr,G)$ terminates with a success
 if and only if $G$ follows from
$\Pscr$ in intuitionistic linear logic. Further, it respects the declarative
reading of all the logical connectives including $\add$.
\end{theorem}

\section{Some Examples }\label{sec:modules}

Let us first consider the relation $f(X,Y)$ specified by two rules:

\begin{numberedlist}

\item if $X < 2$, then $Y = 0$.

\item if $X \geq 2$, then $Y = 3$.

\end{numberedlist}

The two  conditions are mutually exclusive which is expressed by using the cut in traditional logic
programming as shown below:

\[ f(X,0) :-\ X < 2, !. \]
\[ f(X,3) :-\ X \geq 2. \]

\noindent Using cut, we can specify mutually exclusive goals, but  cuts
affect the declarative meaning of the program. Our language makes it possible to formulate mutually exclusive goals
 through the choice-disjunctive goals as shown below:

\begin{exmple}
$f(X,Y)$ ${\rm :-}$ \> \hspace{8em} $(X\geq 2 \otimes Y = 3)\ \add $\\
 \> \hspace{8em} $(X < 2 \otimes Y = 0)$\\
\end{exmple}
\noindent The new program, equipped with $\add$-goals, is more readable
than the original version with cuts, while preserving the same efficiency.
A similar example is provided by the 
following ``max'' program that finds the larger of two numbers.  

\begin{exmple}
$max(X,Y,Max)$ ${\rm :-}$ \> \hspace{9em} $(X\geq Y \otimes Max = X)\ \add $\\
 \> \hspace{9em} $(X < Y \otimes Max = Y)$\\
\end{exmple}
\noindent These two goals in the body of the above clause are mutually exclusive. Hence, only one of these two goals
can  succeed.  For example, consider a goal $max(3,9,Max)$.
Solving this goal  has the effect of choosing and executing the second goal $(3 < 9) \otimes Max = 9$, producing
the result $Max = 9$.

As another example, we consider  the relation $member(X,L)$ for establishing whether $X$ is in the list $L$.
A typical Prolog  definition of $member(X,L)$ is shown below:

\begin{exmple}
$member(X,[Y|L])$ ${\rm :-}$ \> \hspace{9em} $(Y = X)\ \lor member(X,L)$\\
\end{exmple}
\noindent This definition is nondeterministic in the sense that it can find any occurrence of $X$.
Our language in Section 2 makes it possible to change $member$ to be deterministic and more efficient: 
only one occurrence can be
found. An example of  this is provided by the following
 program.

\begin{exmple}
$member(X,[Y|L])$ ${\rm :-}$ \> \hspace{9em} $(Y = X)\ \add 
                                              member(X,L)$\\
\end{exmple}

\section{Conclusion}\label{sec:conc}

In this paper, we have considered an extension to Prolog with  
choice-disjunctive  goals. This extension allows goals of 
the form  $G_0 \add  G_1$  where $G_0, G_1$ are goals.
These goals are 
 particularly useful for replacing the cut in Prolog, making Prolog
more concise and more readable.

In the near future, we plan 
to  investigate the connection between 
\muprolog and Japaridze's Computability Logic(CL)\cite{Jap03,Jap08}.
CL is a new semantic platform for reinterpreting logic
as a theory of tasks. Formulas in CL stand for instructions
that can carry out some  tasks. We plan to investigate whether  our operational semantics is sound and complete
with respect to  the semantics of CL. 

\section{Acknowledgements}

This research was supported by Dong-A University Research Fund.


\end{document}